\documentclass[a4paper]{article}

\usepackage{INTERSPEECH2020}
\usepackage{makecell}
\usepackage{multirow}
\begin{document}
\title{Deformable TDNN with adaptive receptive fields for speech recognition}
\name{Keyu An, Yi Zhang, Zhijian Ou$^{\dagger}$\thanks{$\dagger$ Corresponding author.}}
\address{
  Speech Processing and Machine Intelligence (SPMI) Lab, Tsinghua University, China}
\email{aky19@mails.tsinghua.edu.cn, zhangyi18@mails.tsinghua.edu.cn, ozj@tsinghua.edu.cn}
\maketitle
\ninept
\begin{abstract}
Time Delay Neural Networks (TDNNs) are widely used in both DNN-HMM based hybrid speech recognition systems and recent end-to-end systems.  Nevertheless, the receptive fields of TDNNs are limited and fixed, which is not desirable for tasks like speech recognition, where the temporal dynamics of speech are varied and affected by many factors. In this paper, we propose to use deformable TDNNs for adaptive temporal dynamics modeling in end-to-end speech recognition. Inspired by deformable ConvNets, deformable TDNNs augment the temporal sampling locations with additional offsets and  learn the offsets automatically based on the ASR criterion, without additional supervision. Experiments show that deformable TDNNs obtain state-of-the-art results on WSJ benchmarks (1.42\%/3.45\% WER on WSJ eval92/dev93 respectively), outperforming standard TDNNs significantly. Furthermore, we propose the latency control mechanism for deformable TDNNs, which enables deformable TDNNs to do streaming ASR without accuracy degradation.
\end{abstract}

\section{Introduction}
Time Delay Neural Networks (TDNNs), also known as one-dimensional Convolutional Neural Networks (1-D CNNs), are widely used in both DNN-HMM based hybrid speech recognition systems \cite{Peddinti2015ATD, TDNN_F} and recent end-to-end systems \cite{EE-LF-MMI,CTC-CRF-NAS}. For speech recognition tasks, TDNNs and their variants have several nice properties. First, due to the pure feed-forward nature, the computation in training TDNNs can be easily paralleled. The training of TDNNs is much faster than recurrent neural networks, especially when using GPUs.  Second, it is more convenient to deploy TDNNs for online streaming speech recognition.  In TDNNs, future context information, which is essential for accurate acoustic modeling, is provided by temporal convolution over the future frames. The low-latency of TDNNs is determined by the kernel size and the number of layers, which can be easily implemented. In contrast, neural architectures like bidirectional LSTMs, where the future context is provided by the whole-utterance backward LSTMs, and self-attention encoders, which need the entire utterance as input, are not naturally suited to low-latency speech recognition, although there have some modifications and progress \cite{Transformer-Based-Online,SAA}.

\begin{figure}[ht]
  \centering
  \includegraphics[width=\linewidth]{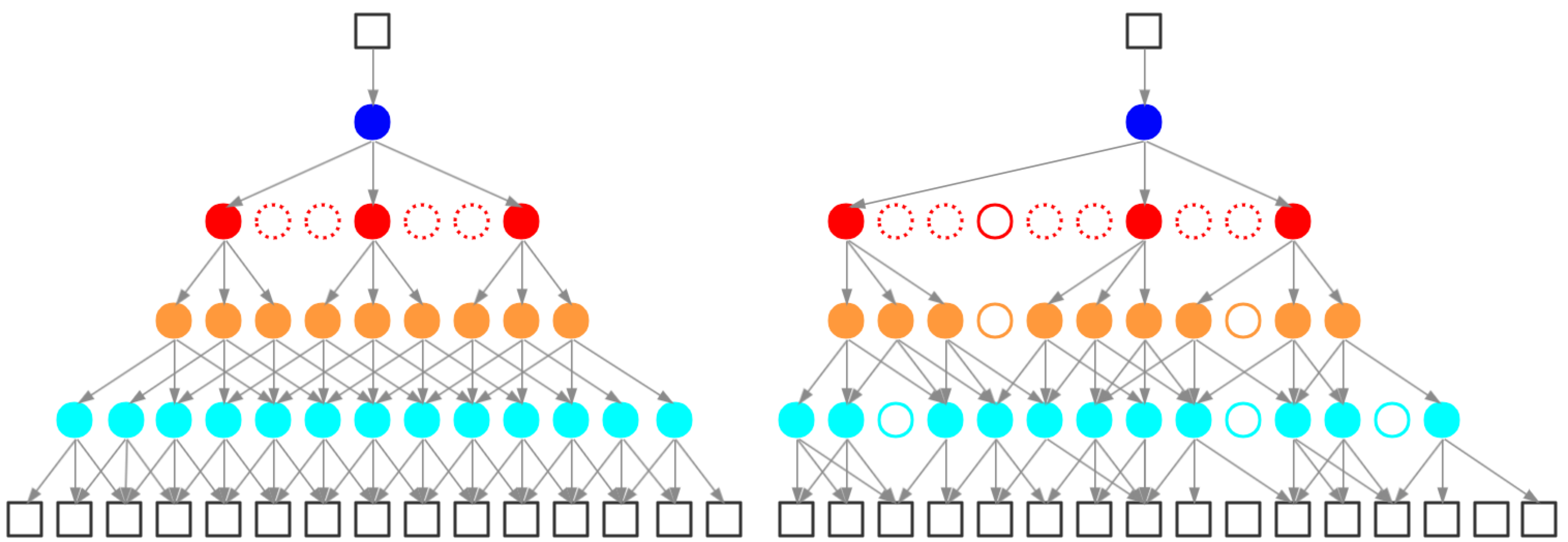}
  \caption{ Illustration of the fixed receptive field in standard TDNNs (left) and the adaptive receptive field in deformable TDNNs (right). Dotted circles denote the dilation between kernel elements.}
  \label{fig:comparison}
  \vspace{-0.25cm}
\end{figure}

Nevertheless, due to the fixed geometric structures, standard TDNNs are inherently limited in modeling the complex transformations of the input features (MFCC, filter bank, and PLP, etc.) needed for speech recognition.
To be specific, the patterns of receptive fields \footnote{In TDNNs, a window of input features around the current position are used to produce the output for the current neural unit. By convention we call the window of features as the receptive field of the current unit.} of all neural units in a TDNN layer are the same, which means that for all neural units in a TDNN layer, input features with \textbf{fixed} positions and 
windows are used to predict linguistic states such as phonemes or characters (Figure \ref{fig:comparison}, left). 
This is undesirable for tasks like speech recognition, where the temporal windows of speech frames covering different linguistic states are varied and affected by many factors, such as the type of phonemes (vowel v.s. consonant), the gender and the pronunciation habit of the speaker, co-articulation, and so on \cite{Multi-Stream-Self-Attention}. To address this, multi-stream models  \cite{Multi-Stream-Self-Attention, Multistream-CNN} propose to accommodate diverse temporal resolutions in multiple streams to achieve robustness in
temporal dynamics modeling. However, in the above-mentioned multi-stream models, the receptive fields for all neural units are pre-designed by human experts, thus can not automatically adapt to the variations in the input. Moreover, multi-stream models significantly increase the number of model parameters.

This paper presents deformable TDNNs (Figure  \ref{fig:comparison}, right) as a neural network architecture for adaptive receptive fields in speech recognition. The idea of deformable TDNN is inspired by deformable convolutional networks (deformable ConvNets)  \cite{DCN}, which was proposed to enhance CNNs’ capability in modeling geometric transformations for sophisticated vision tasks such as object detection and semantic segmentation. Basically, a deformable TDNN can be seen as a 1-D version of deformable ConvNets. Similar to deformable ConvNets, deformable TDNNs augment the temporal sampling locations with additional offsets. The offsets are predicted by an additional TDNN conditioned on the input, thus enabling temporal modeling in a self-adaptive manner. The parameters of the additional TDNN are optimized using the loss for the target task, without additional supervision. Our extensive experiments show that deformable TDNNs improve over standard TDNNs significantly and consistently, without the cost of expensive training time, large overhead over model parameters, and complex hyper parameter designs.

It is important to note that similar modules have been used in action segmentation \cite{TDRN} and sign language recognition \cite{TDC} in videos. The contributions of this work are: 
1)  We demonstrate that adaptive temporal modeling is crucial for speech recognition by comparing the performances of several neural architectures with different receptive fields.
2) We show that deformable TDNNs can replace standard TDNNs easily in existing neural architectures, and produce remarkable performance gains, especially when the input is perturbed in the time direction.
3) We introduce the latency control mechanism for deformable TDNNs, which enables deformable TDNNs to do streaming ASR in practical scenarios.

The rest of the paper is organized as follows. Section 2 outlines related work, especially focusing on the comparison of receptive fields used by different neural architectures. Section 3 describes the deformable TDNNs in details. Section 4 describes the experiment setup. Section 5 presents and analyzes the results primarily on the WSJ task. In Section 6, we discuss the receptive fields of deformable TDNNs, the distribution of the offsets, and the latency control mechanism. Section 7 presents the conclusions and the future work.
We will release the code in the CAT toolkit \cite{CAT} upon the acceptance of the paper.

\section{Related work}
\subsection{Receptive fields of neural architectures}
The receptive fields of different neural architectures are different by their nature. In convolutional neural networks (e.g. 2-D CNNs and TDNNs), the receptive field is associated with the kernel sizes and dilation rates  \cite{ Multi-Stream-Self-Attention,Multistream-CNN}. The FSMN model \cite{fsmn} is another feedforward architecture. In FSMN, context information is encoded into a fixed-size memory block, and the receptive field is defined by the number of historical items looking back to the past, and the size of the lookahead window into the future.
In self-attention layers \cite{transformer}, the output at each time step depends on the entire input feature map, with attention weights conditioned on the input as well.
For RNNs, the receptive field can be seen as the input feature at and before the current time step, due to the recurrent manner. Similarly, the output of bidirectional RNNs relies on the entire input feature map. In LSTMs \cite{LSTM} and GRUs \cite{GRU}, two variants of RNN, the use of forget gate allows a dynamically changing contextual window over all of the input history. Extensive results show that LSTMs and GRUs improve significantly over vanilla RNNs.

The effect of neural architecture's receptive field has been explored in many previous works. In  \cite{Peddinti2015ATD}, the input time steps, at which activations are computed at each layer, are manually designed to model the temporal dependencies. In  \cite{CTC-CRF-NAS}, the hyper-parameters, which control the input contexts for each TDNN layer (the kernel size and dilation), are selected by Neural Architecture Search (NAS). The above-mentioned two works show that the selection of input contexts, either by human experts or by NAS, has a remarkable effect on the model performance.
Time-restricted self-attention mechanism  \cite{Time-Restricted-Self-Attention} restricts the input context within a local region, and suggests that too wide or too narrow a context might degrade the results. \cite{WindowedAttention} proposes a novel fully-trainable windowed attention mechanism to restrict attention to a small input time span. In windowed attention, the window size is learned instead of pre-set by rules.
For recurrent neural networks, Soft forgetting  \cite{SF} and Contextualized Soft forgetting \cite{CAT} propose to unroll the BLSTM network over chunks (without or with overlaps) instead of the whole utterance, thus limiting the dependencies between input and output within the chunks. Results show that such restrictions bring considerable recognition accuracy improvements in addition to realizing streaming, presumably because chunk-based models alleviate overfitting.

\subsection{Multi-resolution networks}
There have been efforts in considering the varieties of temporal patterns and achieving robust acoustic modeling by accommodating diverse temporal information in multiple streams.  In multistream CNNs  \cite{Multistream-CNN}, each stream stacks TDNN-F layers whose kernel has a unique, stream-specific dilation rate. In multi-stride self-attention  \cite{Multi-Stride-Self-Attention}, each group of heads in self-attention process speech frames with a unique stride over neighboring frames. Nevertheless, the hyper-parameters (the dilation rate and the stride) in previous multi-resolution networks are pre-designed by human experts and can not adapt to the input.  Moreover, multi-resolution networks typically result in a significant increase in the number of parameters.

\subsection{Deformable neural networks}
Deformable convolution \cite{DCN} was initially introduced in computer vision tasks to enhance the transformation modeling capability of CNNs. In action segmentation \cite{TDRN} and sign language recognition  \cite{TDC}, where the inputs are organized as a sequence, deformable convolutions were used to model the variations in temporal contexts. Recently, deformable CNNs were applied to the speaker verification task to deal with the irregular structure in feature maps \cite{DCN_SV}. To the best of our knowledge, there was no related work on applying deformable neural networks in speech recognition.

\section{Deformable TDNNs}
\subsection{Method}
A standard TDNN \cite{TDNN} consists of two steps:  1) sampling
with fixed 1-D grid $\mathcal{R}$ over the input feature map $x$; 2)
summation of sampled values weighted by $w$. The 1-D grid
defines the receptive field size and dilation. For example, a TDNN layer with kernel size 3 and dilation 2 can be described as: 
\begin{equation} \label{tdnn} 
y(t_0) = \sum_{t_n \in \mathcal{R}} w(t_n) * x(t_0+t_n)
\end{equation}
where $\mathcal{R}$ = \{-2,0,2\}. 

Similar to deformable convolutions, in a deformable TDNN, the regular 1-D grid $\mathcal{R}$ is augmented with offsets  $\{ \Delta t_n | n = 1, ..., N \}$ \footnote{In 1-D convolution, $N$ equals to the kernel size. 
For notation brevity, we omit the dependence of $\Delta t_n$ on $t_0$.
}, where  $N=|\mathcal{R}|$.
Eq. \ref{tdnn}  becomes
\begin{equation} \label{deformable tdnn} 
y(t_0) = \sum_{t_n \in \mathcal{R}} w(t_n) * x(t_0+t_n+\Delta t_n )
\end{equation}
where $ t = t_0+t_n+\Delta t_n $  can be fractional, as $\Delta t_n$ is typically fractional. Thus, $x(t)$ is calculated via linear interpolation as
\begin{equation}
x(t) = x(\lfloor t \rfloor)*(\lfloor t \rfloor +1 - t) + x(\lfloor t \rfloor + 1)*(t - \lfloor t \rfloor )
\end{equation}
where $\lfloor   \rfloor$ denotes the floor function, defined as giving the greatest integer less than or equal to the input.

\begin{figure}[ht]
  \centering
  \includegraphics[width=\linewidth]{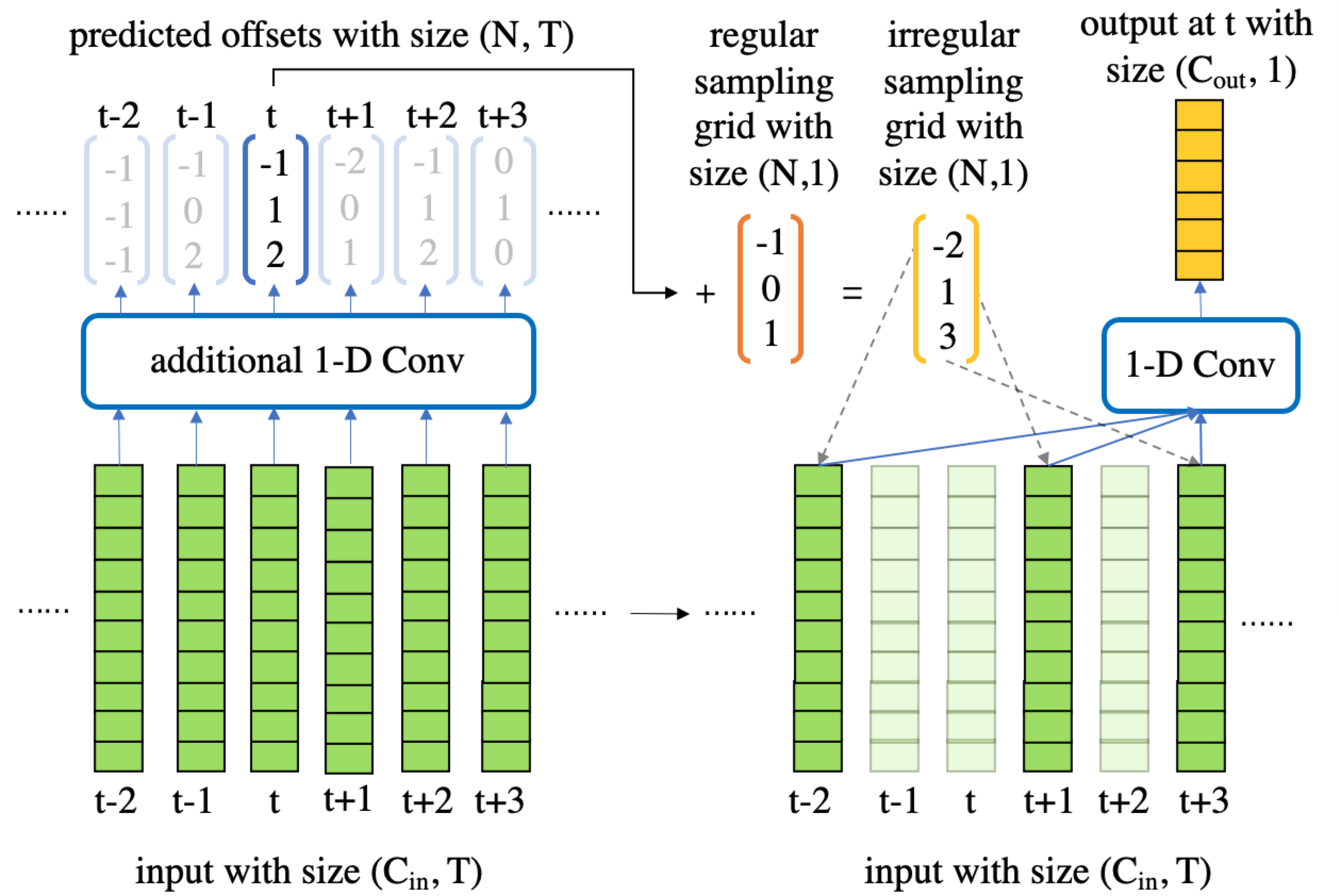}
  \caption{ Illustration of a deformable TDNN with kernel size 3. For the sake of simplicity, we set the predicted offsets to be integers in this figure.}
  \label{fig:deformTDNN}
  \vspace{-0.25cm}
\end{figure}

Consider an input sequence of size ($C_{in},T$), where $C_{in}$ is the dimension of the feature vector, and $T$ is the length of the feature sequence (e.g. the number of frames). Then, the offsets of the deformable TDNN are of size ($N,T$) (Figure \ref{fig:deformTDNN}), predicted by an additional 1-D Conv with kernel size $N^{\prime}$.  The number of additional parameters introduced by the offset predicting TDNN is small ($N * C_{in} * N^{\prime}$), compared with the parameters of regular 1-D Conv ($C_{out} * C_{in} * N$), as $N^{\prime}$ is much smaller than $C_{out}$ typically \footnote{In our experiment, $N^{\prime} = 5$ and $C_{out} = 640$ by default .}.

As illustrated in Figure \ref{fig:deformTDNN}, the offsets
determine the kernel pattern of the deformable TDNN, and the kernel pattern is optimized at the same time as the kernel weights, without additional supervision. The deformation of the kernel pattern depends on the input features,  which enables adaptive adjustment according to the distortions of the input. In this sense, deformable TDNN shares similar high-level spirit with self-attention network \cite{relationship}.
\subsection{Comparison with deformable ConvNets}
Given a feature map (the input spectrogram or the hidden layers in a TDNN) with $T$ time steps, we can view it as an image where the time axis is horizontal and the frequency axis is vertical. When computing the output of a deformable TDNN at each time step, all points along the vertical axis share the same offsets. Thus, the spatial resolution of the offset is much smaller than the input feature map, which leads to fewer additional parameters and computing costs.  Moreover, the offsets are constrained to have only 1 direction (horizontal). In other words, the offsets along the vertical axis are prohibited, as our motivation is to enhance the model's capability in modeling temporal dynamics.
This is different from deformable ConvNets, where the offset fields have the same spatial resolution as the input feature map, and the offsets have 2 directions (e.g., for deformable ConvNets, an input of size ($H,W$) would yield output offsets of size ($2N,H,W$)).

\section{Experiment Setup}
For running the experiments, we use the CTC-CRF-based ASR Toolkit, CAT \cite{CAT}. CAT is based on the recently proposed CTC-CRF~\cite{CRF_IC20}.
Similar to CTC, CTC-CRF uses a hidden state sequence to obtain the alignment between the label sequence and input feature sequence, and the hidden state sequence is mapped to a unique label sequence by removing consecutive repetitive labels and blanks. In CTC-CRF, the posteriori of the hidden state sequence is defined by a conditional random field (CRF), and the potential function of the CRF is defined as the sum of node potential and edge potential. The definition of node potential is the same as CTC, and the edge potential is realized by an n-gram LM of labels.  By incorporating the probability of the sentence into the potential, the conditional independence between the hidden states in CTC is naturally avoided. It has been shown that CTC-CRF outperforms regular CTC consistently on a wide range of benchmarks, and is on par with other state-of-the-art end-to-end models \cite{CTC-CRF-NAS,CAT,CRF_IC20}.

We choose the neural architecture proposed in  \cite{CTC-CRF-NAS} as the baseline acoustic model, which is a 7-layer standard TDNN with hyper-parameters (the kernel size and dilation for each layer) chosen by NAS. The configuration of the baseline model is listed in the Table \ref{baseline}. It is a strong enough baseline (Table \ref{wsj}), and we evaluate the effect of deformable TDNNs by replacing the standard TDNN layers with deformable TDNN layers. In a deformable TDNN layer, the additional offset prediction network is a 1-D Convolution layer. We set the kernel size to 5 by default, and we initialize the parameters of the offset predicting network with 0.
\begin{table}[ht]
\vspace{-0.25cm}
	\centering
	\caption{Configuration of the baseline TDNN model obtained by NAS.}
		\vspace{-0.25cm}
	\scalebox{1.0}{
	\begin{tabular}{cccccccc}
		\toprule
		\textbf{layer index} & \textbf{1} & \textbf{2}  &  \textbf{3} & \textbf{4} & \textbf{5} & \textbf{6} & \textbf{7}\\
        \midrule
	    kernel size  & 5  & 5 & 5&  3 &5  & 5 & 5 \\
	    dilation & 1 & 2 & 2 & 1& 1 & 1 & 2\\
        stride  & 1  & 1 & 1&  3 &1  & 1 & 1  \\
		\bottomrule
	\end{tabular}}
\vspace{-0.25cm}
	\label{baseline}
\end{table}

Experiments are conducted on 80-hour WSJ dataset. Input features are 40-dimension fbank with delta and delta-delta features (120 dimensions in total). The features are augmented by 3-fold speed perturbation, and are mean and variance normalized.

\section{Experiment}
Table \ref{usage} shows the effect of deformable TDNNs on WSJ. By replacing standard TDNNs in the last 1, 2, 3 layers with deformable TDNNs, recognition accuracy increases remarkably and consistently on both dev93 and eval92, with only a small overhead over model parameters.  This indicates that the significant performance improvement is from the capability of modeling temporal dynamics, rather than increasing model parameters. As the evaluation accuracy stops improving when more than 2 deformable TDNN layers are used, we use deformable TDNNs for the last two layers in the remaining experiments.
\begin{table}[ht]
\vspace{-0.25cm}
	\centering
	\caption{WER (\%) results of using deformable TDNN in the last 1, 2, 3 TDNN layers. The numbers in parentheses denote the relative performance improvements via the use of deformable TDNN layers.}
		\vspace{-0.25cm}
	\scalebox{0.85}{
	\begin{tabular}{cccc}
		\toprule
		\makecell[c]{\textbf{usage of deformable}\\ \textbf{  TDNN (\# layers) }} & \textbf{\# parameters}  &  \textbf{dev93} & \textbf{eval92} \\
        \midrule
	    none (0, baseline) \cite{CTC-CRF-NAS} & 11.90 M & 5.68 & 2.77 \\
		\midrule
	    layer 7 (1)  & 11.92 M & \textbf{5.37} (5.5\%) & 2.64 (4.7\%)\\
        layer 6,7 (2)  & 11.93 M & 5.48 (3.5\%) & \textbf{2.50} (9.7\%) \\
        layer 5,6,7 (3)  & 11.95 M & 5.54 (2.5\%) & 2.55 (7.9\%) \\
		\bottomrule
	\end{tabular}}
\vspace{-0.25cm}
	\label{usage}
\end{table}

To evaluate the robustness of deformable TDNNs, we compare the performance of standard TDNNs and deformable TDNNs when  the input spectrum is warped along the time axis during testing.  Specifically, we employ the time warping operation proposed in SpecAugment \cite{specaug}, where a random point on the feature map is to be warped either to the left or right by a distance $w$ chosen from a uniform distribution from 0 to $W$ (the time warp parameter) along the time axis. To reduce the randomness, we conduct every comparison experiment 10 times with different random seeds. As shown in table \ref{time_warp}, deformable TDNNs perform consistently better at adapting to the variations in the input.
\begin{table}[ht]
\vspace{-0.25cm}
	\centering
	\caption{WER (\%) results on WSJ eval92 when the input feature is warped either to the left or right by a distance $w$ chosen from a uniform distribution $\mathcal{U}(0,W)$}
		\vspace{-0.25cm}
	\scalebox{0.85}{
	\begin{tabular}{cccc}
		\toprule
		\textbf{Time warp parameter $W$} & \textbf{80}  &  \textbf{160} & \textbf{240} \\
        \midrule
	    Standard TDNN & 3.10 & 6.41 & 11.05 \\
		\midrule
	    Deformable TDNN  & 2.98 & 5.93 & 10.31 \\
		\bottomrule
	\end{tabular}}
\vspace{-0.25cm}
	\label{time_warp}
\end{table}

We compare CTC-CRF with deformable TDNNs and other models in Table \ref{wsj}. From Table \ref{wsj}, we find that: 1) CTC-CRF trained with deformable TDNNs improve significantly over CTC-CRF with other neural architectures, including popular architectures such as BLSTM and self-attention encoders. 2) Two popular methods in speech recognition, SpecAugment  \cite{specaug} and RNNLM lattice rescoring  \cite{rnnrescore} further improve the recognition accuracy. 3) With deformable TDNNs,  mono-phone CTC-CRF performs significantly better than other end-to-end/hybrid models. Notably, deformable TDNN based CTC-CRF obtains the lowest word error rate (WER) of 1.42\%/3.45\% on WSJ eval92/dev93 among all published ASR results, to the best of our knowledge.
\begin{table}[ht]
\vspace{-0.25cm}
	\centering{
	\caption{WER ($\%$) results of CTC-CRF with deformable TDNNs v.s. other models on WSJ.}
		\vspace{-0.25cm}
	\scalebox{0.85}{
	\begin{tabular}{ccccc}
		\toprule
	\multicolumn{2}{c}{\textbf{Model}}    & \textbf{Unit} & \textbf{dev93} & \textbf{eval92} \\
	    \midrule
	\multicolumn{2}{c}{LF-MMI \cite{SS-LF-MMI}} & mono-phone & 6.0 & 3.0 \\
    \multicolumn{2}{c}{LF-MMI \cite{SS-LF-MMI}} & bi-phone & 5.3 & 2.7 \\
	\multicolumn{2}{c}{E2E-LF-MMI \cite{SS-LF-MMI}} & mono-phone & 6.3 & 3.1 \\
	\multicolumn{2}{c}{E2E-LF-MMI \cite{SS-LF-MMI}} & bi-phone & 6.0 & 3.0 \\
	\multicolumn{2}{c}{ESPRESSO  \cite{Wang2019Espresso}} & subword & 5.9 & 3.4 \\
    \multicolumn{2}{c}{ESPnet \cite{espnet} VGGBLSTM AM} & char & 8.8 & 5.3 \\
    \multicolumn{2}{c}{ESPnet \cite{espnet} transformer AM} & char & 6.6 & 4.6 \\
    \multicolumn{2}{c}{ConvAM + ConvLM  \cite{Zeghidour2018fully}} & char & 6.8 & 3.5 \\
    \multicolumn{2}{c}{Jasper + transformerLM  \cite{jasper}} & char & 9.3 & 6.9 \\
		\midrule
	\multicolumn{1}{c}{\multirow{3}{*}{CTC-CRF}}& BLSTM \cite{CRF_IC20} & mono-phone & 6.23 & 3.79 \\
	~ & VGGBLSTM \cite{CAT} & mono-phone & 5.7 & 3.2 \\
	~ & self-attention $^{\rm 1}$ & mono-phone & 6.24 & 3.12 \\
	~ & TDNN-NAS \cite{CTC-CRF-NAS} & mono-phone & 5.68 & 2.77 \\
	\midrule
    \multicolumn{1}{c}{\multirow{3}{*}{CTC-CRF}}& Deformable TDNN & mono-phone & 5.48 & 2.50 \\
    ~ & + SpecAugment & mono-phone &5.38 & 2.46 \\
    ~ & + RNN LM $^{\rm 2}$ & mono-phone & 3.45 & 1.42 \\
		\bottomrule
	\end{tabular}
	\label{wsj}}
	\footnotesize{$^{\rm 1}$Obtained by our implementation. $^{\rm 2}$ The neural network for RNN LM is a 3-layer TDNN interleaved with LSTM layers.}}
	\vspace{-0.25cm}
\end{table}
\section{Discussion}
\subsection{Receptive fields}
\begin{figure}[ht]
\vspace{-0.25cm}
  \centering
  \includegraphics[width=0.45\linewidth,height=0.225\linewidth]{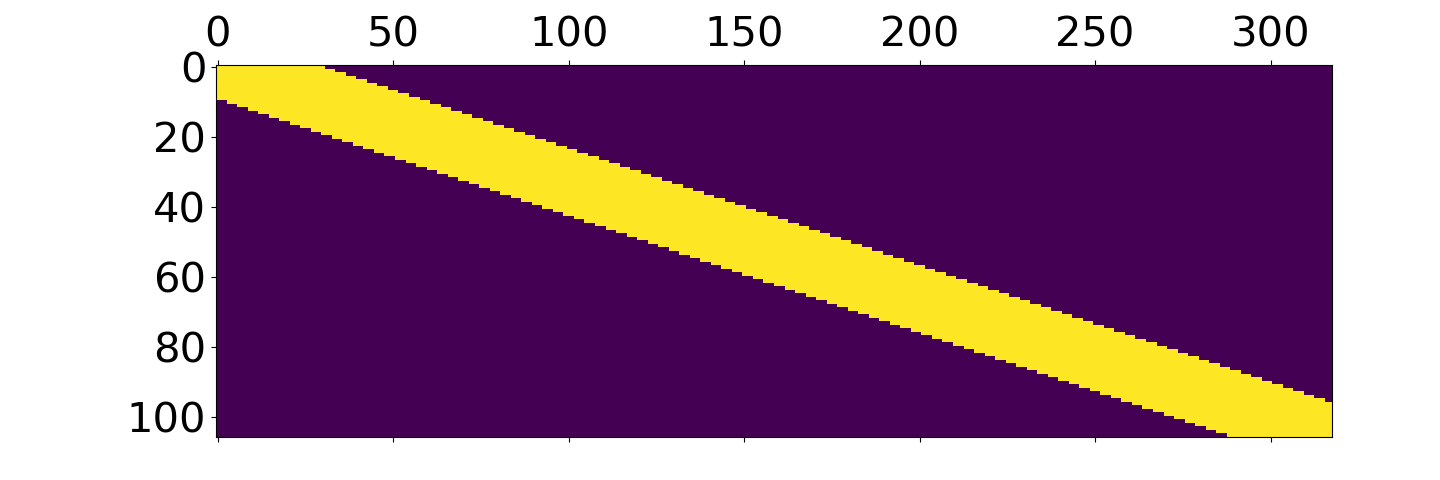}
  \includegraphics[width=0.45\linewidth,height=0.225\linewidth]{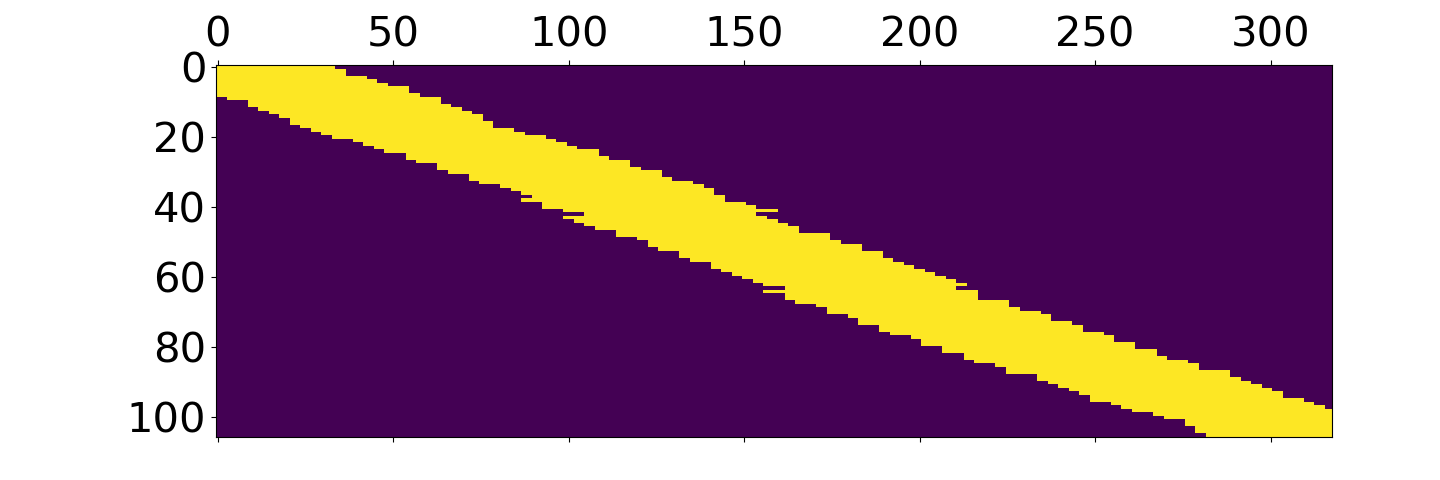}
  \caption{The receptive fields of standard TDNNs (left) and deformable TDNNs (right). For every grid [i, j], yellow means output[i] depends on input[j], while purple means output[i] is independent of input[j]. The input is three times as long as the output because of the convolution stride. The results are for the utterance (444c040x) in the WSJ eval92 set. }
  \label{fig:receptive_field}
  \vspace{-0.25cm}
\end{figure}

The receptive fields of standard TDNNs and deformable TDNNs are shown in Figure \ref{fig:receptive_field}. From Figure \ref{fig:receptive_field}, we observe that: 1) The receptive fields of deformable TDNNs follow the monotonic order in general, which is desired for speech recognition, where the feature inputs and corresponding outputs generally proceed in the same order. 2) Compared with standard TDNNs, the receptive fields of deformable TDNNs are more flexible. The receptive field size for different outputs varies, which is somewhat similar to the alignment patterns of attention models \cite{hybrid_CTC_Atten}.
\subsection{Distribution of the offsets}
\begin{figure}[ht]
\vspace{-0.25cm}
  \centering
  \includegraphics[width=\linewidth,height=0.5\linewidth]{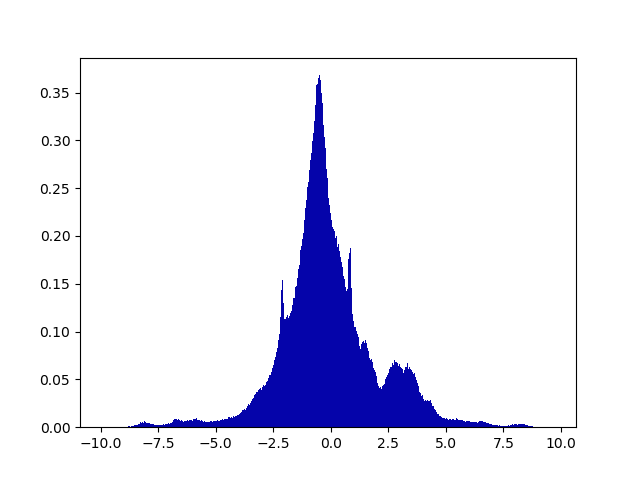}
  \caption{Distribution of predicted offsets on WSJ eval92 sets.}
  \label{fig:offsets}
  \vspace{-0.25cm}
\end{figure}

The distribution of predicted offsets on WSJ eval92 sets is shown in Figure \ref{fig:offsets}. We observe that: 1) Although most of the offsets are close to zeros, a small but significant number of offsets are big in magnitude, which could produce noticeable deformations of the kernel patterns. 2) Most predicted offsets are less than 0, indicating that the model tends to pay more ``attention'' to  previous inputs to produce the current prediction.
\subsection{Latency control mechanism}
One of the advantages of TDNNs is their predictable and relatively low latency. However, in deformable TDNNs, the additional offsets may lead to unpredictable dependency on future frames, causing uncontrollable latency. A direct solution is to clip the offset during testing, e.g., set the predicted offset to 0 if it is greater than 0. Thus, the usage of deformable TDNN layers will not cause more latency than standard TDNNs. However, in our experiments, this approach brings an intolerable decline in recognition accuracy (Table \ref{latency}). To address it, we apply the clip mechanism in both training and inference. After applying latency controlled training, deformable TDNNs are able to do streaming speech recognition without the degradation of recognition accuracy.
\begin{table}[ht]
\vspace{-0.25cm}
	\centering
	\caption{WER (\%) results on WSJ when using different strategies for latency control in training and testing.}
		\vspace{-0.25cm}
	\scalebox{0.85}{
	\begin{tabular}{cccc}
		\toprule
		\textbf{Training} & \textbf{Testing}   &  \textbf{dev93} & \textbf{eval92} \\
        \midrule
	    No constrains  & No constrains & 5.48 & 2.50\\
        No constrains  & latency controlled & 7.70 & 4.38 \\
        latency controlled  & latency controlled  & 5.54  & 2.60\\
		\bottomrule
	\end{tabular}}
\vspace{-0.25cm}
	\label{latency}
\end{table}
\section{Conclusion}
This paper introduces deformable TDNNs for end-to-end speech recognition, featured by adaptability to input variations, superior results, and capability of streaming speech recognition.  Moreover, deformable TDNNs can replace TDNNs easily in existing neural architectures and can be trained in an end-to-end manner without extra supervision. Future works include evaluating the effectiveness of deformable TDNNs on large-scale datasets and exploring the combination with other neural architectures (e.g., LSTMs, self-attention layers).

\bibliographystyle{IEEEtran}

\bibliography{template}

\end{document}